\documentclass[11pt,twoside]{article}
\usepackage{asp2006_hotcool}
\usepackage{graphicx}
\begin{document}

\title{\centering The Eclipsing Black Hole X-ray Binary M33 X-7:  \\  \centering  Understanding the Current Properties}  

\author{Francesca Valsecchi, Bart Willems, Tassos Fragos, and Vicky Kalogera} 
\affil{Northwestern University, Department of Physics and Astronomy, 2145 Sheridan Road, Evanston, IL 60208, USA}   
\begin{abstract} 
We explore the formation and evolution of the black hole X-ray binary system M33 X-7. In particular, we examine whether accounting for systematic errors in the stellar parameters inherent to  single star models, as well as the uncertainty in the distance to M33, can explain the discrepancy between the observed and expected luminosity of the $\sim70\,$M$_\odot$ companion star.
Our analysis assumes no prior interactions between the companion star and the black hole progenitor.Ê
We use four different stellar evolution codes, mo\-di\-fied to include a variety of current stellar wind prescriptions. For the models satisfying the observational constraints on the donor star's effective temperature and luminosity,Êwe recalculate the black hole mass, the orbital separation, and the mean X-ray luminosity.Ê Our best model, satisfying simultaneously all observational constraints except the observationally inferred companion mass, consists of a ${\sim13\,}$M$_\odot$ black hole and a ${\sim54\,}$M$_\odot$ companion star. 
We conclude that a star with the observed mass and luminosity can not be explained via single star evolution models, and that a prior interaction between the companion star and the black hole progenitor should be taken into account.
\end{abstract}

\pagestyle{myheadings}
\section{Introduction}   
M33 X-7  is the first stellar-mass black hole (BH) discovered in an eclipsing X-ray binary system (XRB). \citet{Orosz2007} constrained the mass to be ${\sim15.65\,}$M$_\odot$ for the BH and ${\sim70\,}$M$_\odot$ for the O-star companion. The orbital period ($P$) of ${\sim3.45\,}$days was determined by \citet{Pietsch2006}. The distance has been measured to be between ${\sim 840\,}$kpc \citep{Orosz2007} and ${\sim 960\,}$kpc \citep{Bonanos2006}. Table \ref{params} and \ref{dist} summarize other  system's parameters relevant to our analysis.
\begin{table}[ ! ht]
\caption{Observed parameters for M33 X-7.}
\label{params}
\begin{center}
{\small
\begin{tabular}{lrclr}
\noalign{\smallskip}
\tableline
\noalign{\smallskip}
$M\rm_{don}$(M$_\odot$) & 70.0 $\pm$ 6.9 &\phantom{a} & $M\rm_{BH}$(M$_\odot$) &15.65 $\pm$ 1.45 \\
$R\rm_{don}$(R$_\odot$)  &19.6 $\pm$ 0.9 & & $a$(R$_\odot$)  &42.4 $\pm$ 1.5 \\
$P$(days)  &3.45301 $\pm$ 0.00002 & & $i$(\deg) &74.6 $\pm$ 1.0 \\
$f(M\rm_{BH}$)(M$_\odot$)  &0.46 $\pm$ 0.08 && $Z$(Z$_\odot$)$^*$  &5\% to 40\%   \\
Spectral Type & O7 III to O8 III && $T$$\rm_{eff}$(K)   &35000 $\pm$ 1000  \\
\noalign{\smallskip}
\tableline
\end{tabular} \\
\footnotetext{}{* Orosz and collaborators, private communication 2008}
}
\end{center}
\end{table}

\begin{table}[ ht]
\caption{Luminosity and X-ray luminosity of M33 X-7 for the distances adopted by \citet{Orosz2007} and \citet{Bonanos2006}. The mean value for the X-ray luminosity according to \citet{Orosz2007} has been calculated from \citet{Liu2008}.}
\label{dist}
\begin{center}
{\small
\begin{tabular}{lrclr }
\tableline
\noalign{\smallskip}
\multicolumn{5}{c}{\citet{Orosz2007}}{\citet{Bonanos2006}} \\
\hline
\noalign{\smallskip}
$d$(kpc) & 840 $\pm$ 20 &\phantom{a} & $d$(kpc) &  964 $\pm$ 54\\ 
$\log(L\rm_{don}$/L$_\odot$)& 5.72 $\pm$ 0.07 &&  $\log(L\rm_{don}$/L$_\odot$)& 5.84 $\pm$ 0.09 \\ 
 $L\rm_{X}$(erg/s)& (2.01 $\pm$ 0.48)$\times10^{38}$ & & $L\rm_{X}$(erg/s)& (2.65 $\pm$ 0.70)$\times10^{38}$ \\ 
\noalign{\smallskip}
\tableline
\end{tabular}
}
\end{center}
\end{table}
It is challenging to make such a tight system with such a massive BH through standard BH-XRB formation channels. The current idea is that the BH currently accretes mass from the wind of the O-star companion. Mass transfer through Roche-lobe overflow (RLO) is excluded by the extreme mass ratio ($q$ $\sim\,$4.5) since RLO would be dynamically unstable and lead to a rapid merger of the two stars. \citet{Orosz2007} also reported that the companion star is less luminous than what is expected from single star models. Because of the instability of mass transfer, this discrepancy cannot be explained by a past RLO phase from the O-star to the BH (mass transfer on the thermal time scale of the mass donor in particular is known to produce underluminous donor stars).

\section{Modelling the System}
The goal of our analysis is to explore whether the observed system parameters, in particular the underluminous nature of the BH companion, can be explained via single star models when systematic errors in the star's parameters inherent to stellar evolution codes, as well as uncertainties in the distance to M33, are taken into account.
We evolve a range of single star models accounting for mass loss through stellar winds, and extract the mass ($M\rm_{don}$) and radius ($R\rm_{don}$) when the model matches the observed effective temperature ($T\rm_{eff}$) and luminosity ($L\rm_{don}$). We consider both the luminosity reported by \citet{Orosz2007} for a distance to M33 of $\sim840\,$kpc and a luminosity rescaled to the distance to M33 of $\sim964\,$kpc reported by \citet{Bonanos2006} (Table \ref{dist}). Given the mass of the secondary when the model matches the observed $T\rm_{eff}$ and $L\rm_{don}$  we recalculate the mass of the BH ($M\rm_{BH}$) from the observed mass function ($f(M\rm_{BH}$)) and inclination ($i$). Next we calculate the orbital separation ($a$), the secondary's Roche-lobe radius, and the X-ray luminosity ($L\rm_X$). We restrict our investigation to those stars that are currently not filling their Roche-lobe.

\subsection{Uncertainties in Modeling Massive Stars}  
\begin{figure}[!ht]
\begin{center}
\resizebox{8cm}{!}{\includegraphics{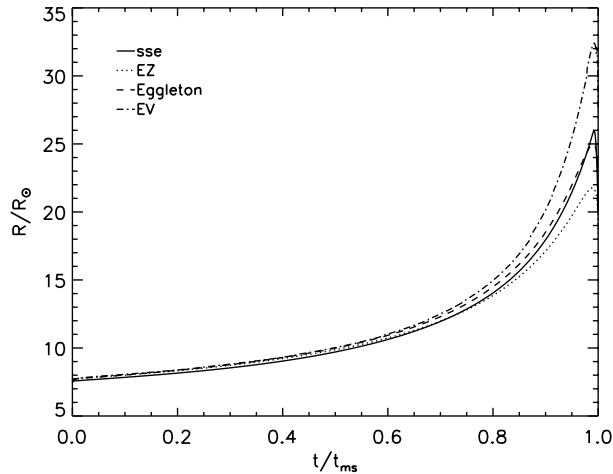}}
\end{center}
\caption{Radial evolution of a $50\,$M$_\odot$ star as a function of time in units of the main-sequence life time for four different stellar evolution codes: SSE \citep{Hurley2000}, EZ \citep{Paxton2004}, Eggleton \citep{Egg2002} and EV (\citeauthor{Nata}\citeyear{Nata}; \citeauthor{Nata2}\citeyear{Nata2}). At the end of the Main Sequence the systematic uncertainty in the radius can reach $5\,$R$_\odot$. } 
\label{RvsTms} 
\end{figure}

The main uncertainties in the single star evolution models are the wind mass-loss rates and systematic errors in radius, luminosity and effective temperature inherent to stellar evolution codes (see Figure \ref{RvsTms} for an example). 
We therefore use four different codes modified to include an up-to-date spectrum of stellar wind prescriptions (\citeauthor{Hurley2000} \citeyear{Hurley2000}; \citeauthor{Vink2001} \citeyear{Vink2001}, and references therein). We then take into account the difference in the calculated radii, effective temperatures and luminosities for every considered single star model. To account for the uncertainty in the observed metallicity ($Z$), we consider models with metallicities ranging from $Z$ = $0.05\,$Z${_\odot}$ to $Z$ = $0.5\,$Z${_\odot}$, but in the present paper we present results only for $Z=0.2\,$Z${_\odot}$. 

\subsection{The Mass of the O-Star From Single Stars Models at $Z = 0.004$}
\begin{figure}[!ht]
\begin{center}
\resizebox{8cm}{!}{\includegraphics{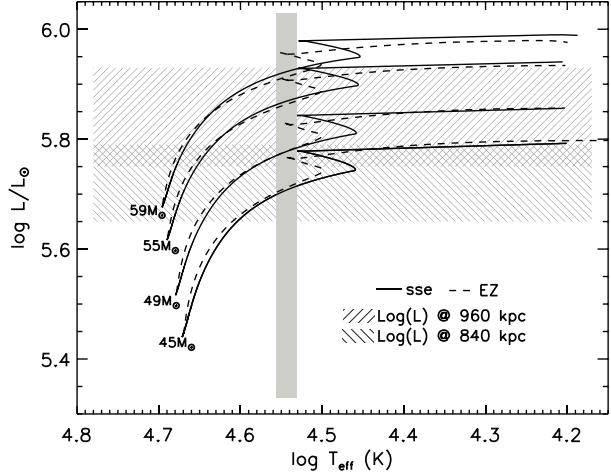}}
\end{center}
\caption{HR diagram for different initial ZAMS secondary masses and metallicity $Z$ = 0.004 calculated with the EZ and SSE stellar evolution codes. The grey area represents the observational constraints on $T\rm_{eff}$, while the hatched areas represent the observational constraints on $L_{\rm don}$ for distances of 840 and $960\,$kpc. } 
\label{HRdiag} 
\end{figure}

The current observational constraints on the companion's radius, luminosity and effective temperature, narrow the range of possible masses that can be considered.
Figure \ref{HRdiag} shows HR diagram for different initial secondary masses calculated with the EZ and SSE stellar evolution codes. The SSE tracks cross the observed range of $T\rm_{eff}$ one time. The EZ tracks cross the observed $T\rm_{eff}$ three times, with the third crossing being in the Hertzsprung gap (HG). Given the low probability of observing a star in the HG we do not consider the third crossing any further. Using a distance to M33 of $\sim 840\,$kpc, the observed luminosity and effective temperature constraints can be satisfied by masses up to $49\,$M$_\odot$ for the 1${^{st}}$ crossing found with both SSE and EZ, while the constraints can only be satisfied by masses up to $46\,$M$_\odot$ for the 2${^{nd}}$ crossing found with EZ. Assuming a distance  of  $\sim 960\,$ kpc , the constraints can be satisfied by masses up to $59\,$M$_\odot$ for the 1${^{st}}$ crossing found with both SSE and EZ, while the constraints can only be satisfied by masses up to $56\,$M$_\odot$ for the 2${^{nd}}$ crossing found with EZ.
 
\section{Results and Conclusions}

\begin{figure}[!ht]
\begin{center}
\resizebox{6.5cm}{!}{\includegraphics{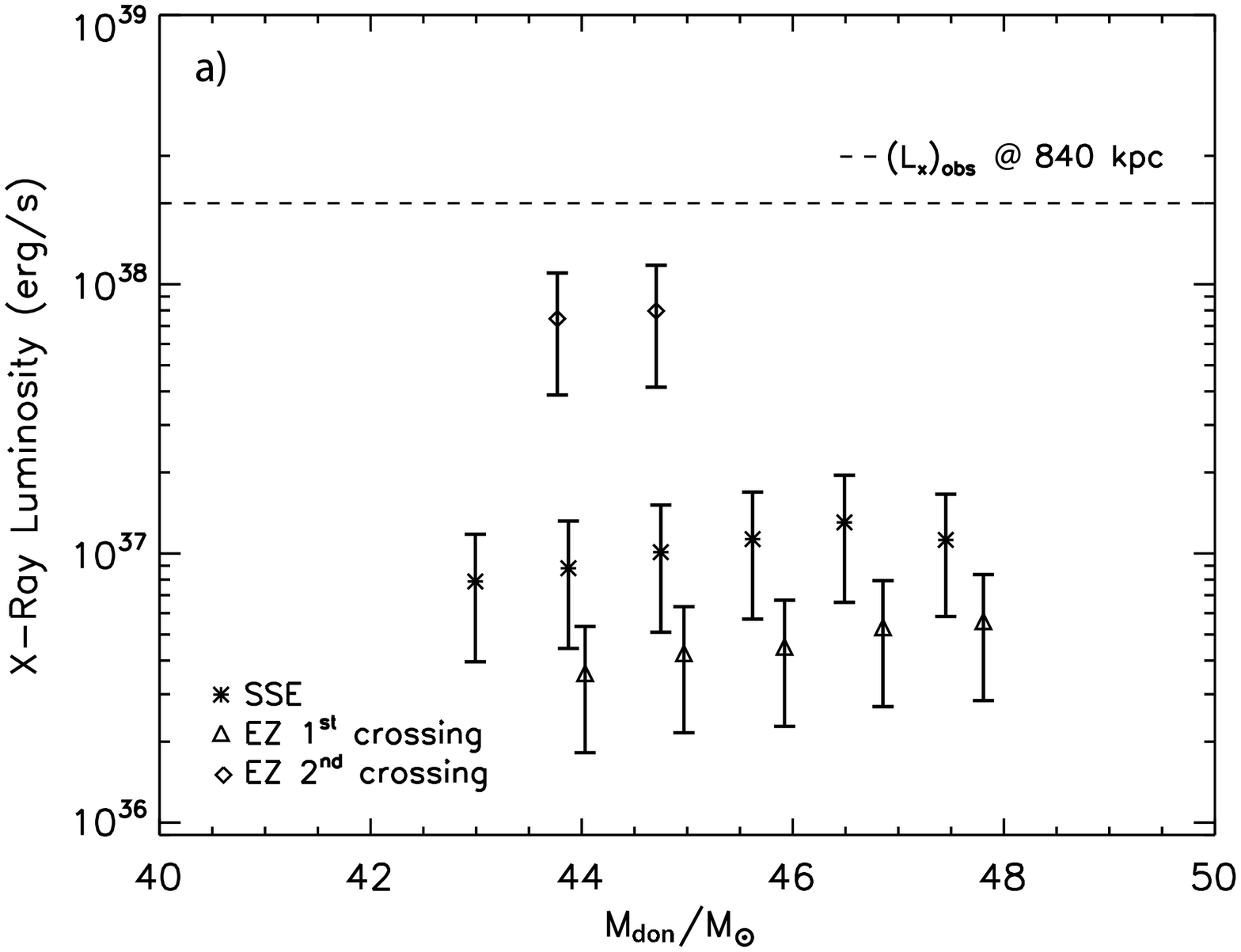}} \resizebox{6.5cm}{!} {\includegraphics{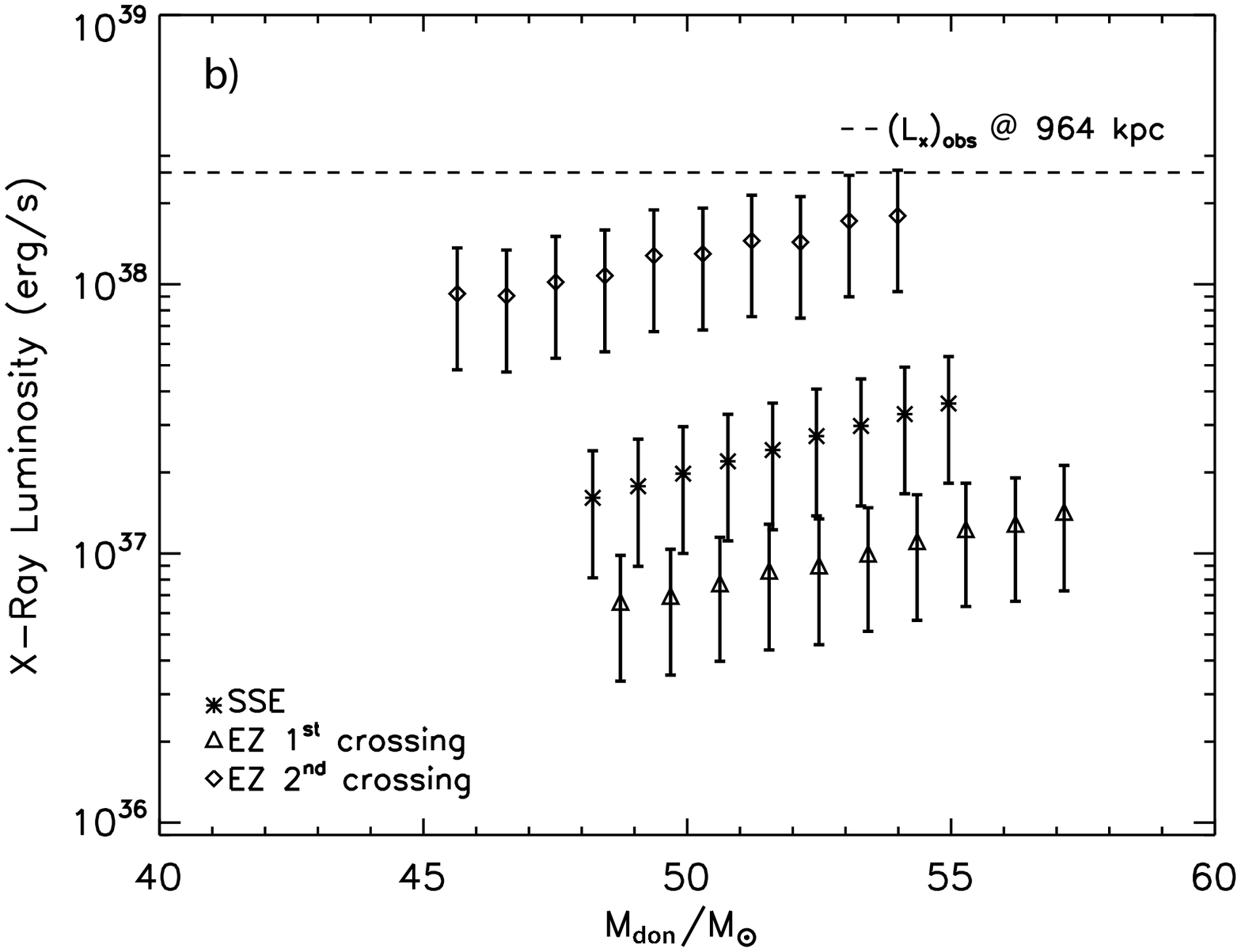}}
\end{center}
\caption{X-ray luminosity as a function of present day donor mass for $Z$ = 0.004, according to single stars evolution models, for distances of $\sim840\,$ (panel a) and $\sim960\,$kpc (panel b). Accounting for the derived upper limits on the initial secondary mass for the 2$^{nd}$ crossing of the observed $T\rm_{eff}$ constraints (see Figure \ref{HRdiag}), we could only consider ZAMS masses up to $56\,$M$_\odot$.The dashed horizontal line represents the observed X-ray luminosity.} 
\label{Lx} 
\end{figure}

The main results of our analysis are summarized in Figure \ref{Lx} and in Table \ref{results}
If we assume a distance to M33 of $\sim840\,$kpc, the maximum value for the X-ray luminosity  calculated with EZ is $\sim1.2\times10^{38}\,$erg/s, which is in agreement with the observed value at this distance to within a factor of $\sim1.2$. On the other hand, adopting a distance of $\sim960\,$kpc, the maximum X-ray luminosity calculated with EZ is $\sim2.65\times10^{38}\,$erg/s which is in agreement within the error bars with  the observed X-ray luminosity at this distance. 
As far as the O-star mass is concerned, the highest initial companion masses satisfying the effective temperature and luminosity constraints at  $\sim840\,$kpc are $50\,$M$_\odot$ and $46\,$M$_\odot$  for SSE and EZ respectively. On the other hand,  assuming a distance of $\sim960\,$kpc, we can consider masses up to $59\,$M$_\odot$ and $56\,$M$_\odot$ for SSE and EZ, respectively.
Therefore, a zero-age main-sequence star (ZAMS) of $56\,$M$_\odot$ is the most massive single star model that satisfies simultaneously the observational constraints on the effective temperature, donor luminosity and X-ray luminosity. The calculated masses for this model for the O-star and the BH are  $\sim54\,$M$_\odot$ and $\sim13\,$M$_\odot$, respectively.
We conclude that a companion star of $70\,$M$_\odot$ with the observed luminosity ($\log(L\rm_{don}$/L$_\odot$)$\sim 5.72$) can not be explained via single star evolution models. Accounting for systematic errors in radius, luminosity and effective temperature inherent to stellar evolution codes, as well as uncertainties in the distance to M33 does not solve this discrepancy.

\begin{table}[ ! ht]
\label{results}
\caption{Donor  mass, BH  mass,  orbital separation, and maximum X-ray luminosity calculated for the highest initial donor mass satisfying the $T\rm_{eff}$ and $L\rm_{don}$ constraints ($50\,$M$_\odot$ and $46\,$M$_\odot$  for SSE and EZ respectively for a distance of  $\sim 840\,$ kpc,  $59\,$M$_\odot$ and $56\,$M$_\odot$ for SSE and EZ respectively for a distance of $\sim 960\,$ kpc). }
\begin{center}
{\small
\begin{tabular}{lcccc}
\tableline
\noalign{\smallskip}
Code & SSE & EZ  & SSE & EZ\\
$d$(kpc)-Crossing & 840 - 1$^{st}$ & 840 - 2$^{nd}$  & 964 - 1$^{st}$ & 964 - 2$^{nd}$\\
$M\rm_{don}$ initial (M$_\odot$) & 50 & 46 &59 & 56 \\
$M\rm_{don}$ @ present (M$_\odot$)  &47.45 $\pm$ 0.03 & 44.71 $\pm$ 0.01 &54.9 $\pm$ 0.2 &53.99 $\pm$ 0.01  \\
$M\rm_{BH}$ @ present (M$_\odot$)  &12.2 $\pm$ 0.8 & 11.8 $\pm$ 0.8 & 13.4 $\pm$ 0.9 & 13.2 $\pm$ 0.2  \\
$a$ @ present (R$_\odot$)  &37.5 $\pm$ 0.2 & 36.9 $\pm$ 0.2 & 39.3 $\pm$ 0.2 & 39.1 $\pm$ 0.2  \\
Max $L\rm_X$ (erg/s)  & $\sim1.66\times10^{37}$ &$\sim1.18\times10^{38}$ & $\sim5.39\times10^{37}$ & $\sim2.65\times10^{38}$  \\
\noalign{\smallskip}
\tableline
\end{tabular}
}
\end{center}
\end{table}

\section{Present and Future Work}
It is important to emphasize that our analysis treats the companion star as a single star, and does not account for any interaction between the companion star and the BH progenitor. 
Since single star models can not explain the observed properties of the O-star, the system likely underwent a mass transfer phase during which the BH progenitor filled its Roche-lobe and transferred mass to the O-star. \citet{Braun1995} shown that, depending on the efficiency of semi-convective mixing, such a mass transfer phase can lead to underluminous secondary stars (the so-called non-rejuvenation scenario).
Therefore, we are currently studying mass transfer sequences to explore this possibility. This analysis will also yield theoretical constraints on the age of the system, the binary component masses, the orbital separation right before the supernova explosion that formed the BH, and the mass lost and any possible natal kick imparted to the BH during the supernova explosion.

\acknowledgements This work was supported by
the NSF CAREER grant AST-0449558 and a
Packard Fellowship in Science \& Engineering to
VK. Numerical simulations were performed on
the HPC cluster available to the Theoretical
Astrophysics Group at Northwestern University
through the NSF MRI grant PHY-0619274
to VK.


\begin{thebibliography}{}
\bibitem[Bonanos et al.(2006)]{Bonanos2006}
Bonanos, A. Z., Stanek, K. Z., Kudritzki, R. P., Macri, L. M., Sasselov, D. D.,Kaluzny, J., Stetson, P. B., Bersier, D., Bresolin, F., Matheson, T., Mochejska, B. J., Przybilla, N., Szentgyorgyi, A. H., Tonry, J., \& Torres, G.  2006, \apj, 652,  313-322
\bibitem[Braun \& Langer (1995)]{Braun1995}
Braun, H., \& Langer, N. 1995, \aap, 297, 483
\bibitem[Eggleton \& Kiseleva-Eggleton (2002)]{Egg2002}
Eggleton, P. P., \& Kiseleva-Eggleton, L. 2002, \apj, 575, 1, 461-473
\bibitem[Hurley et al.(2000)]{Hurley2000}
Hurley, J. R., Pols, O. R., \& Tout, C. A. 2000, \mnras,  315, 3, 543-569
\bibitem[Ivanova et al. (2003)]{Nata2}
Ivanova, N., Belczynski, K., Kalogera, V., Rasio, F. A., \& Taam, R. E. 2003, \apj, 592, 475
\bibitem[Liu et al.(2008)]{Liu2008}
Liu, J., McClintock, J. E., Narayan, R., Davis, S. W., \& Orosz, J. A.  2008, \apj, 679, L37-L40
\bibitem[Orosz et al.(2007)]{Orosz2007}
Orosz, J.A., McClintock, J.E., Narayan, R., Bailyn, C.D., Hartman J.D., Macri, L., Liu, J., Pietsch, W., Remillard, R.A., Shporer, A., \& Mazeh, T. 2007, \nat, 449, 872-875
\bibitem[Paxton (2004)]{Paxton2004}
Paxton, B. 2004, \pasp, 116, 821, 699-701
\bibitem[Pietsch et al.(2006)]{Pietsch2006}
Pietsch, W., Haberl, F., Sasaki, M., Gaetz, T. J., Plucinsky, P. P., Ghavamian, P., Long, K. S., \& Pannuti, T. G. 2006, \apj, 646, 420Ð428
\bibitem[Podsiadlowski et al. (2002b)]{Nata}
Podsiadlowski, P., Rappaport, S., \& Pfahl, E. D. 2002b, \apj, 565, 1107
\bibitem[Vink et al.(2001)]{Vink2001}
Vink, J. S., de Koter, A., \& Lamers, H. J. G. L. M. 2001, \aap, 369, 574-588
\end{thebibliography}
\end{document}